# BOSE-EINSTEIN CONDENSATION IN DISSIPATIVE SYSTEMS FAR FROM THERMAL EQUILIBRIUM

K.Staliunas

Physikalisch Technische Bundesanstalt, 38116 Braunschweig, Germany
tel.: +49-531-5924482, Fax: +49-531-5924423, E-mail: Kestutis.Staliunas@PTB.DE

**Abstract**

It is shown, that Bose-Einstein condensation can occur not only in spatially extended *equilibrium* systems, but also in the systems *far from thermal equilibrium*, which show order-disorder phase transition. The investigation is performed by solving the Complex Ginzburg-Landau equation, a universal model for nonequilibrium order-disorder phase transitions.

Under Bose-Einstein Condensate (BEC) an *equilibrium* state of a *quantum* system is usually understood, such that most of the particles of the system condense into one, the *lowest* energy state of the system [1].

In the preceding paper [2] we showed, that Bose-Einstein condensation may occur not only in quantum systems, but also in classical systems. Essential for the condensation is the coherent behavior of the individual members (particles) in the spatial domain, which is equivalent to autocatalytic behavior in momentum space. In the quantum case this autocatalytic dynamics occurs through the indistinguishability of the individual particles which reduces the probability of random scattering from occupied states. In classical systems such autocatalytic dynamics may occur as well due to nonlinear interaction among the particles.

The present paper deals with another aspect of Bose-Einstein condensation: It is shown that the assumption of thermal equilibrium is not necessary in order to obtain statistical Bose-Einstein distributions. Bose-Einstein distribution can be obtained far from thermal equilibrium as well. One example is the laser. The photons in a laser resonator are far from a thermal equilibrium and the distribution of mode occupations in lasers shows a sharp peak at the lasing (transverse or longitudinal) mode: the photons of the laser radiation condense into one lasing mode. The laser is indeed a system showing an order-disorder transition: below the generation threshold a laser emits incoherent radiation, with exponential (thermal) intensity distribution. Above the generation threshold laser emits a coherent radiation, which is of poissonian intensity distribution.

Analogously to this example of the photon condensation in a laser we search for Bose-Einstein condensation in general systems showing order-disorder phase transitions. Order - disorder phase transitions in spatially extended nonlinear systems are universally described by a complex Ginzburg - Landau equation (CGLE) with a stochastic term:

$$\frac{\partial A}{\partial t} = pA - (1+ic)|A|^2 A + (1+ib)\nabla^2 A + \Gamma(\mathbf{r},t) \tag{1}$$

where $A(\mathbf{r},t)$ is the order parameter defined in n-dimensional space $\mathbf{r}$, and evolving in time $t$. $p$ is the control parameter (the order - disorder transition occurs at $p = 0$). The Laplacian

$\nabla^2 A$ represents the nonlocality in the system, and $\Gamma(\mathbf{r},t)$ is the noise, $\delta$- correlated in space and time, of power (temperature) $T$: $\langle \Gamma(\mathbf{r}_1,t_1) \cdot \Gamma^*(\mathbf{r}_2,t_2) \rangle = 2T \cdot \delta(\mathbf{r}_1 - \mathbf{r}_2)\delta(t_1 - t_2)$

Below the transition threshold ($p<0$) CGLE (1) yields a disordered state: the complex-valued order parameter $A(\mathbf{r},t)$ is essentially noise filtered in space and time, with exponential (thermal) intensity distribution. Above the transition threshold ($p>0$) (1) yields an ordered, or coherent state (or a condensate) in a modulationally stable case, with the order parameter distributed near its mean value $\langle |A|^2 \rangle = p$.

CGLE (1), with complex-valued coefficients, has been derived systematically for many systems showing second order phase transitions in the presence of noise, e.g.: for lasers with spatial degrees of freedom [3], where $A(\mathbf{r},t)$ is proportional to the amplitude of the optical field, and $\Gamma(\mathbf{r},t)$ corresponds to the vacuum or thermal fluctuations; for finite temperature superfluids [4]. The conservative limit of CGLE (the so called Gross-Pitaevskii equation) has been derived for the finite temperature Bose-Einstein condensates [5], where $A(\mathbf{r},t)$ is the wave-function of the condensate, and $\Gamma(\mathbf{r},t)$ corresponds to the fluctuations of thermal bath. The CGLE (1) can be also used to describe the atom laser. The CGLE with real-valued coefficients $b=c=0$, has been systematically derived as the amplitude equation for stripe patterns in nonequilibrium dynamical systems [6], where the amplitude and phase of the order parameter corresponds to the amplitude- and phase modulations of the roll patterns respectively.

The CGLE with real-valued coefficients $b=c=0$, can be written phenomenologically as a normal form, or minimal equation, describing universally nonequilibrium order - disorder phase transitions in the presence of noise [7]: the first two terms $(pA - |A|^2 A)$ approximate in the lowest order a supercritical Hopf bifurcation, and the diffusion term $\nabla^2 A$ describes the simplest possible nonlocality. The complex-valued character of the order parameter $A(\mathbf{r},t)$ is important: every ordered, or coherent state, both in classical physics or in quantum mechanics, is characterized not only by the modulus of the order parameter, but also by its phase. Then, as the real Ginzburg-Landau equation (that with the real-valued order parameter) is the normal form for second order phase transitions between two arbitrary states [8], the CGLE (1) can serve the as the normal form of second order phase transitions between ordered and disordered states of the matter.

It is shown in this letter, that the spatial noise spectra of the CGLE with the real-valued coefficients are of $1/k^2$ - form, where k is the wavenumber of the transverse modes. (For simplicity the case of real-valued coefficients $b=c=0$ is considered in present paper, however the main results are also valid for CGLE with complex-valued coefficients, as commented below.) In this way the statistical distributions of excitations of transverse modes are Bose-Einstein-like. In the long wavelength limit ($k \to 0$) the statistical distributions coincide with the Bose-Einstein distribution [1] derived originally for systems in thermal equilibrium.

For analytical treatment it is assumed that the system is sufficiently far away from the order - disorder phase transition: $p \gg T$. Then the homogeneous component $|A_0| = \sqrt{p}$ is dominating, and one can look for a solution of (1) in the form of a perturbed homogeneous state: $A(\mathbf{r},t) = A_0 + a(\mathbf{r},t)$. Linearizing (1) around the stationary homogeneous solution one obtains:

$$\frac{\partial a}{\partial t} = -p(a + a^*) + \nabla^2 a + \Gamma(\mathbf{r},t) \qquad (2)$$

and its complex conjugate. Diagonalisation of (2): $b_+ = (a+a^*)/\sqrt{2}$ and $b_- = (a-a^*)/\sqrt{2}$, yields:

$$\frac{\partial b_+}{\partial t} = -2pb_+ + \nabla^2 b_+ + \Gamma_+(\mathbf{r},t) \tag{3.a}$$

$$\frac{\partial b_-}{\partial t} = \nabla^2 b_- + \Gamma_-(\mathbf{r},t) \tag{3.b}$$

This shows that:

a) all amplitude fluctuations $b_+$ decay above the phase-transition point with a decay rate $\lambda_+ = -2p - k^2$, where $k$ is the spatial perturbation wavenumber. Asymptotically long-lived amplitude perturbations are possible only at the phase transition point (in a critical state), but never above it;

b) the phase fluctuations $b_-$ decay with a rate $\lambda_- = -k^2$, which means that the long-wavelength perturbations decay asymptotically slowly, with a decay rate approaching zero for $k \to 0$.

From (3) one can calculate spatio-temporal noise spectra, by rewriting (3) in terms of the spatial and temporal Fourier components $b_\pm(\mathbf{r},t) = \int b_\pm(\mathbf{k},\omega)\exp(i\omega t - i\mathbf{k}\mathbf{r})d\omega d\mathbf{k}$: This leads to:

$$b_+(\mathbf{k},\omega) = \frac{\Gamma_+(\mathbf{k},\omega)}{i\omega + \mathbf{k}^2 + 2p} \tag{4.a}$$

$$b_-(\mathbf{k},\omega) = \frac{\Gamma_-(\mathbf{k},\omega)}{i\omega + \mathbf{k}^2} \tag{4.b}$$

The spatio-temporal power spectra are:

$$S_+(\mathbf{k},\omega) = |b_+(\mathbf{k},\omega)|^2 = \frac{|\Gamma_+(\mathbf{k},\omega)|^2}{\omega^2 + (2p+\mathbf{k}^2)^2} \tag{5.a}$$

$$S_-(\mathbf{k},\omega) = |b_-(\mathbf{k},\omega)|^2 = \frac{|\Gamma_-(\mathbf{k},\omega)|^2}{\omega^2 + \mathbf{k}^4} \tag{5.b}$$

for the amplitude and phase fluctuations correspondingly. Assuming $\delta$- correlated noise in space and time, $|\Gamma_\pm(\mathbf{k},\omega)|^2$ are proportional to the temperature $T$ of the random force.

The spatial spectra are obtained by integration (5) over all frequencies $\omega$: $S_{tot}(\mathbf{k}) = S_{amplitude}(\mathbf{k}) + S_{phase}(\mathbf{k}) = \int S_+(\mathbf{k},\omega)d\omega + \int S_-(\mathbf{k},\omega)d\omega$. (The total power spectrum here is the sum of amplitude and phase power spectra, since the spectral components $b_\pm(\mathbf{r},t)$ are mutually uncorrelated, as follows from (4)). The integration is performed for clarity separately for amplitude and phase fluctuations, and yields:

$$S_+(k) = \int_{-\infty}^{\infty} \frac{T}{\omega^2 + (2p+k^2)^2} d\omega = \frac{T\pi}{k^2 + 2p} \tag{6.a}$$

$$S_-(k) = \int_{-\infty}^{\infty} \frac{T}{\omega^2 + k^4} d\omega = \frac{T\pi}{k^2} . \tag{6.b}$$

This means that the spectrum of phase fluctuations is of the form $1/k^2$ (6.b). The spatial spectrum of amplitude fluctuations is Lorentzian: in the short wavelength limit, $|k|^2 \gg 2p$, the amplitude spectrum is equal to the phase spectrum $S_+(k) = S_-(k)$, as follows from (6.a).

In the long wavelength limit $|k|^2 \ll 2p$, the amplitude fluctuation power spectrum saturates to: $S_+(k \approx 0) = T\pi/(2p)$, and is negligibly small compared to the phase fluctuation spectrum.

We note, that if (5) are integrated with respect to spatial wavenumbers $k$, then temporal noise spectra are obtained. This integration was performed in [9], and leads to $1/f^\alpha$ noises, with depending on the dimensionality of the space: $\alpha = 3/2$ for 1D systems, $\alpha = 1$ for 2D systems, and $\alpha = 1/2$ for 3D systems.

Next we checked the spatial spectra calculated analytically (6) by comparing them with those obtained by numerical integration of CGLE (1). We integrate for this purpose numerically the CGLE (with real valued coefficients) in 1, 2, and 3 dimensions, and average in time the occupations of transverse modes. The main problem in the numerical integration (especially in case of higher dimension of space) is that the numerical discretization restricts a range of spatial wavenumbers. In the 2D case, where a grid of (64*64) was used, the range of wavenumbers is less than two decades. For this reason we performed a series of calculations with different sizes of the integration region, and combined the calculated averaged spatial spectra into one plot. The calculations in Fig.1 (1D case) e.g. were performed on a spatial grid of 64 points with five different sizes of the integrating region. (The calculating facilities allowed to calculate on a larger grid than of 64 points in 1D, however we choose such a coarse grid in 1D in order to compare with the results in 2D, and 3D, where grids larger than (64*64) and (64*64*64) respectively, were hardly possible.) The integration region with the size equal to $l$ result in a set of transverse wavenumbers equal to $2\pi n/l$, where $n$ is the index of the discrete transverse wavenumber. (Discrete wavenumbers (modes) occur due to the periodic boundary conditions used.) In this particular 1D case a discrete set of 64 modes with wavenumbers ranging from a minimal value $k_{min} = 2\pi/l$, to the maximal value $k_{max} = 2\pi \cdot 64/l$ occurs. We performed separate calculations with the size of the integration region $l = l_n = 2\pi \cdot 10^{2.5-n/2}$ ($n=1,...,5$). In this way we obtained the spectra combined from partially overlapping pieces, extending in total over more than three decades.

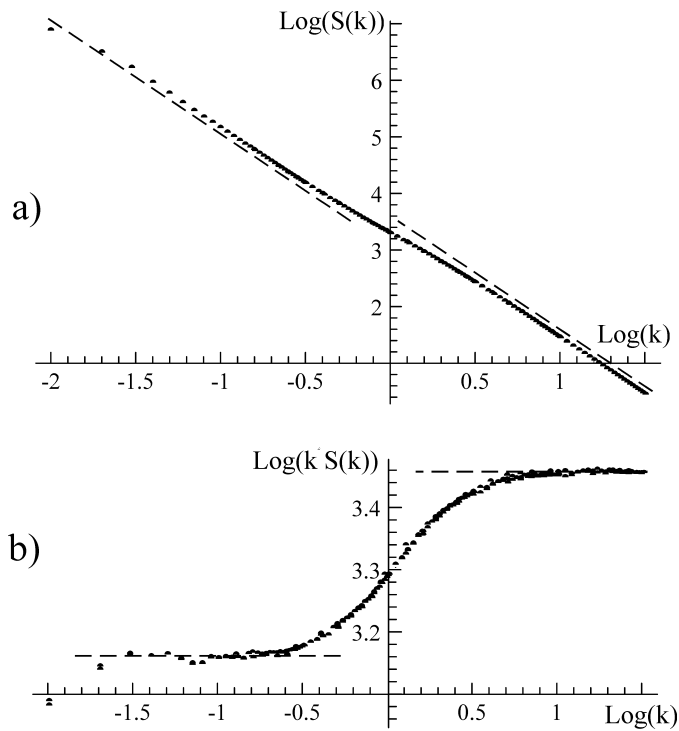

*Fig.1.* The spatial spectrum as obtained by numerical integration of CGLE (1), with real valued coefficients, for 1D. The averaging has been performed over the time of $t = 10^6$. Each point corresponds to averaged intensity of discrete spatial mode. The calculations have been performed with 5 different values of the size of the integration region with different temporal steps:

$l_1 = 2\pi \cdot 10^2$, $\Delta t_1 = 5 \cdot 10^{-2}$;
$l_2 = 2\pi \cdot 10^{1.5}$, $\Delta t_2 = 5 \cdot 10^{-3}$;
$l_3 = 2\pi \cdot 10$, $\Delta t_3 = 5 \cdot 10^{-4}$;
$l_4 = 2\pi \cdot 10^{0.5}$, $\Delta t_3 = 5 \cdot 10^{-5}$;
$l_5 = 2\pi$, $\Delta t_3 = 5 \cdot 10^{-6}$.

These spectra were combined into one plot. The dashed lines correspond to a $1/k^2$ dependence and are for guiding the eye.

The noise spatial power spectrum is shown in Fig.1 for 1D, and in Fig.2 for 2D. The results for 3D show no principal differences with those in 1D and 2D and are not shown.

Fig.1.a and Fig 2.a show the spectra in a log-log scale, where the character of $1/k^2$ is clearly seen. The dashed lines are the lines with $1/k^2$ slope to guide the eye. The $1/k^2$ dependence is good in the limits of long and short wavelengths, however, for the intermediate values of $k$ a discrepancy is observed. This discrepancy is most clearly seen from Fig.1.b and Fig.2.b, where the normalized spectra $k^2 S(k)$ are plotted. The "kink" at intermediate values of $k$ joins the spectra in the limits of long and short wavelengths which are both of the same slope, but of different intensities. As follows from (6) for long wavelengths ($k \to 0$) the amplitude fluctuation spectrum is negligible compared to the phase fluctuation spectrum $S_+(k) << S_-(k)$, and the total spectrum is: $S(k) = S_+(k)$. For short wavelengths ($k \to \infty$) the amplitude fluctuation spectrum is equal to the phase fluctuation spectrum $S_+(k) = S_-(k)$, and the total spectrum is: $S(k) = 2S_+(k)$. The numerical results shown in Fig.1 and Fig.2. confirm this result.

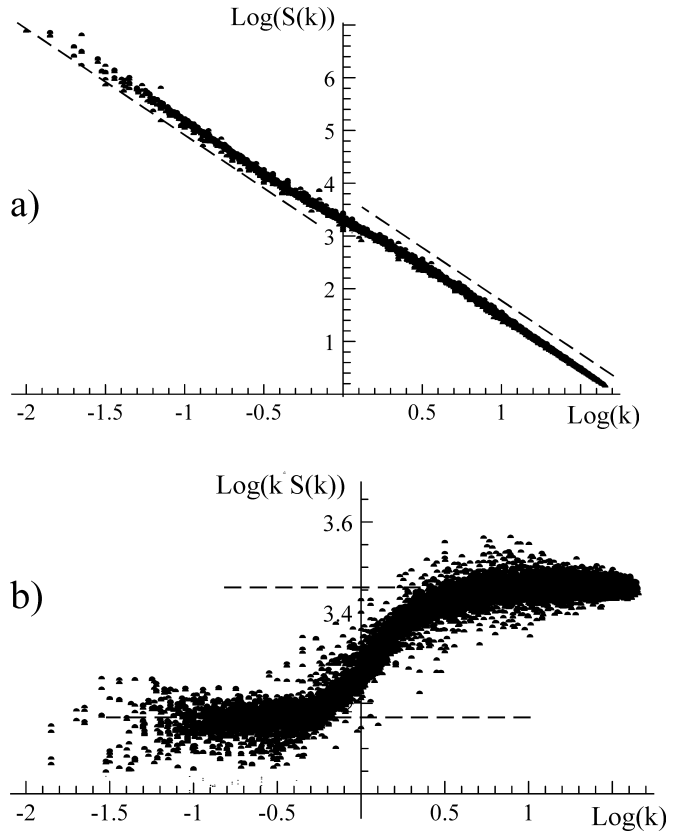

*Fig.2.* The spatial spectrum as obtained by numerical integration of CGLE (1) for 2D. The averaging has been performed over the time of $t = 10^4$. Everything else as in Fig.1.

One more reasons why we combined the spectra from pieces calculated separately was the finite size of the temporal step used in our split-step numerical technique. Indeed, in order to obtain the correct spectra in the long wavelength limit the integration is time-consuming. The long waves are very slow, and the characteristic build-up time for long waves is of order of $\tau_{build} \approx 1/k^2$, as seen from (4,5), and diverges for $k \to 0$. One has to average for very long time to obtain the correct statistics for the long waves. On the other hand, the characteristic build-up times for short wavelengths become very small, since the same relation $\tau_{build} \approx 1/k^2$ holds. Here, in order to obtain a correct statistics of mode occupation one has to correspondingly decrease the size of the temporal step for $k \to \infty$. We thus come to the conclusion, that one can never obtain the analytically predicted (correct) $1/k^2$ statistical distribution in a single numerical run, with finite temporal steps (with limited time resolution). The spectrum calculated with a fixed temporal step is shown in Fig.3. In log-log representation (Fig.3.a) a sharp decrease of occupation of the large wavenumbers occurs. In representation of logarithm of spectral density vs. $k^2$ (Fig.3.b) a straight line indicating an exponential decrease is obtained for large wavenumbers. The spectrum shown in Fig.3, curiously enough, is thus precisely a Bose-Einstein distribution, decaying with the power law for long wavelengths: $S(k \to 0) \propto k^{-2}$, and exponentially for short wavelengths: $S(k \to \infty) \propto \exp(-k^2)$.

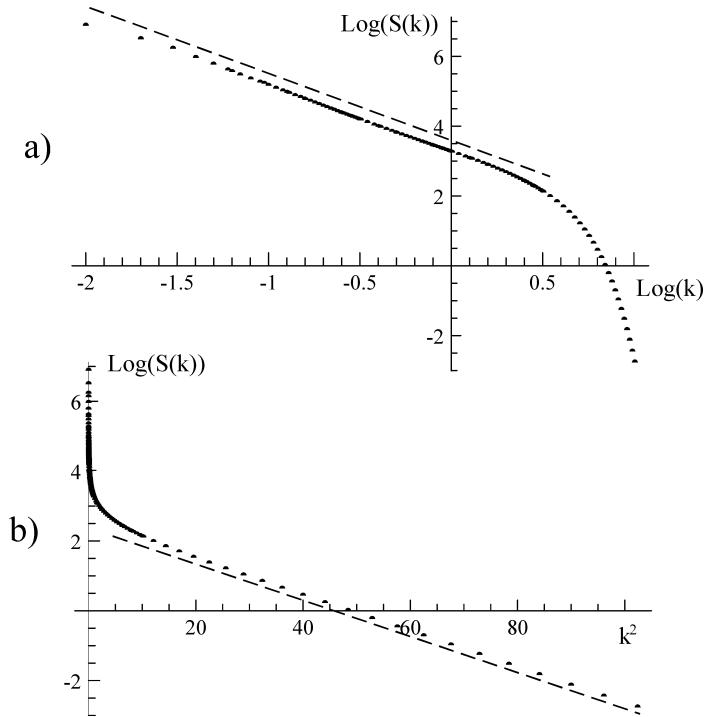

*Fig.3.* The spatial spectrum as obtained by numerical integration of CGLE (1) for 1D for fixed temporal step of $\Delta t = 5 \cdot 10^{-2}$, but combined from four calculations with different size of integration region. The averaging has been performed over the time of $t = 10^6$.
a) shows the spectrum in log-log representation and the dashed line corresponds to $1/k^2$ dependence.
b) shows spectrum in single log representation and the dashed line corresponds to $\exp(-k^2)$ dependence.

We note, that the linear stability analysis does not lead to the exact Bose-Einstein distribution found numerically with finite temporal steps. The finite temporal step $\Delta t$ is equivalent to a particular cut-off frequency $\omega_{max}$ of the temporal spectrum: $\omega_{max} = 2\pi/\Delta t$. In order to account for this finite temporal resolution the integration of (6) should be performed not over the all frequencies, but over $[-\omega_{max}, +\omega_{max}]$. This integration, however, leads to a power law decay $S(k \to \infty) \propto k^{-4}$ for short wavenumbers, and not to the expected exponential decay. We have no explanation for this discrepancy between the analytical and numerical results.

We performed a series of numerical calculations varying the size of the temporal step, in order to interpolate the spectra in the total range of the spatial frequencies. The result is:

$$S(k) = \frac{T\pi C/\omega_{max}}{\exp(k^2 C/\omega_{max}) - 1}. \tag{7}$$

Here $C$ is a constant of order one. (7) reproduces correctly the numerically obtained spectra in both asymptotics of $k \to 0$ and $k \to \infty$. For the intermediate values of the wavelengths $k^2/\omega_{max} \approx 1$ a transition between power law and exponential decay is predicted by (7) exactly as it found in the numerical calculations. In this way numerical results show that the spatial spectrum in the case of limited temporal resolution of the system showing an order-disorder transitions (modeled by CGLE) coincide precisely with the Bose-Einstein distribution, whereas the spectrum in the case of unlimited temporal resolution follows the power law.

The spatial fluctuation spectra were calculated for the CGLE with real-valued coefficients $b = c = 0$. However the results can be directly extended to the case of a CGLE with complex-valued coefficients, at least the in modulationally stable range: $1 + bc > 0$. The eigenvalues for the linearized equation (the analog of (2)) are: $\lambda_+ = -2p - k^2(1-bc)$ and $\lambda_- = -k^2(1+bc)$ for amplitude and phase perturbations, in the long wave limit $k^2 \ll 1$. This generates linear Langevin equations similar to (3), and eventually leads to the same $1/k^2$ spatial spectra. This allows to obtain the Bose-Einstein distributions not only in predominantly dissipative systems $b, c \ll 1$, but to the generalize our result to general case of spatially extended systems showing order-disorder phase transitions.

In the accompanying paper [10] we show, that Bose-Einstein-like distributions in nonequilibrium spatially extended systems may occur not only into the lowest state of the system ($k = 0$), but also into higher, excited modes ($k \neq 0$).

We acknowledge discussions with C.O.Weiss, A.Berzanskis and M.Lewenstein. This work has been supported by Sonderforschungs Bereich 407.

**References**

1. S.N.Bose, Z.Phys. **26** (1924) 178
2. K.Staliunas, Bose-Einstein Condensation in Classical Systems, e-print: http://xxx.lanl.gov/abs/cond-mat/0001347 2000.
3. R.Graham, and H.Haken, Zeit. für Physik **337** (1970) 31.
4. V.L.Ginzburg, and L.D.Landau, Zh.Exp.Theor.Fiz. **20** (1950) 1064, V.L.Ginzburg, and L.P.Pitaevskii, Sov.Phys.JETP **7** (1958) 858.
5. E.P.Gross, Nuovo Cimento **20** (1961) 454; L.P.Pitaevski,Zh.Exp.Teor.Fiz. **40** (1961) 646 [Sov.Phys.JETP **13** (1961) 451].
6. A.C.Newell, and J.A.Whitehead, J.Fluid Mech. **38** (1969) 279; L.A.Segel, J.Fluid Mech. **38** (1969) 203.
7. The relations between Langevin CGLE and the second order phase transition has been discussed in: H.Haken, Synergetics. An Introduction, (Springer Verlag, Berlin, Heidelberg, New York, 1979).
8. L.D.Landau, and E.M.Lifschitz: Course of Theoretical Physics, Vol. 6 and 9 (Pergamon, London, New York 1959). R.Brout, Phase Transitions (Benjamin, New York 1965).
9. K.Staliunas, 1/f Noise in Spatially Extended Systems with Order - Disorder Phase Transitions, submitted to Phys. Rev. Letters 1999; also e-print: http://xxx.lanl.gov/abs/patt-sol/9912004 1999.
10. K.Staliunas, Bose-Einstein Condensation into Higher, Excited States, to appear in preprint server (2000).